\documentclass[twocolumn,aps,prapp,superscriptaddress]{revtex4-1}  
\usepackage{graphicx}  
\usepackage{dcolumn}   
\usepackage{bm}        
\usepackage{amssymb}   
\usepackage{hyperref}
\usepackage{amsmath}
\usepackage{mathtools}

\newcommand{\citeasnoun}[1]{Ref.~\citenum{#1}}
\makeatletter
\newcommand*{\rom}[1]{\expandafter\@slowromancap\romannumeral #1@}
\makeatother

\newcommand{\SII}[1]{Appendix-#1}

\newcommand{\citeSII}[2]{[\citenum{#1}, Appendix-#2]}
\DeclareMathOperator*{\argmax}{arg\,max}
\renewcommand{\vec}[1]{\mathbf{#1}}
\DeclareMathOperator*{\asin}{asin}

\begin{document}

\title{From solar cells to ocean buoys:\\ Wide-bandwidth limits to absorption by metaparticle arrays}

\author{Mohammed Benzaouia} \affiliation{Department of Electrical Engineering and Computer Science, MIT, Cambridge, MA 02139, USA}
\author{Grgur Toki\'c} \affiliation{Department of Mechanical Engineering, MIT, Cambridge, MA 02139, USA}
\author{Owen D. Miller} \affiliation{Department of Applied Physics and Energy Sciences Institute, Yale University, New Haven, CT 06511, USA}
\author{Dick K. P. Yue} \affiliation{Department of Mechanical Engineering, MIT, Cambridge, MA 02139, USA}
\author{Steven G. Johnson} \affiliation{Department of Mathematics, MIT, Cambridge, MA 02139, USA}

\begin{abstract}
 In this paper, we develop an approximate wide-bandwidth upper bound to the absorption enhancement in arrays of metaparticles, applicable to general wave-scattering problems and motivated here by ocean-buoy energy extraction. We show that general limits, including the well-known Yablonovitch result in solar cells, arise from reciprocity conditions. The use of reciprocity in the stochastic regime leads us to a corrected diffusion model from which we derive our main result: an analytical prediction of optimal array absorption that closely matches exact simulations for both random and optimized arrays under angle/frequency averaging. This result also enables us to propose and quantify approaches to increase performance through careful particle design and/or using external reflectors. We show in particular that the use of membranes on the water's surface allows substantial enhancement. 
\end{abstract}
\maketitle

\section{Introduction.} 
One of the most influential theoretical results for solar-cell design has been the ray-optical Yablonovitch limit \cite{yablonovitch1982statistical,green2002lambertian,yu2010fundamental,yu2011angular,sheng2011optimization,wang2012absorption,callahan2012solar,ganapati2014light}, which provides a bound to how much surface texturing can enhance the performance of an absorbing film averaged over a broad bandwidth and angular range. In this paper, we obtain approximate broad-band/angle absorption limits for a case in which the traditional Yablonovitch result is not useful: dilute arrays of \textquotedblleft metaparticles\textquotedblright (synthetic absorbers/scatterers). Known limits bound the absorption at every wavelength \cite{wolgamot2012interaction,buddhiraju2017theory}, but they tend to be loose when considering large bandwidths since coherent effects average out \cite{yu2010fundamentalopt, sheng2011optimization}. Here, we find limits on the absorption for arrays of particles that can be described by the radiative-transfer equation (RTE) \cite{ishimaru1978wave, tsang2000scattering}. In particular, we show that an isotropic diffusive regime is \emph{optimal} for maximizing absorption. This allows us both to obtain analytical upper bounds (Eqs.~\ref{eq6}, \ref{eq1}) and identify the ideal operating regime of absorbing metaparticle arrays.

\begin{figure}
\includegraphics[width=0.5\textwidth]{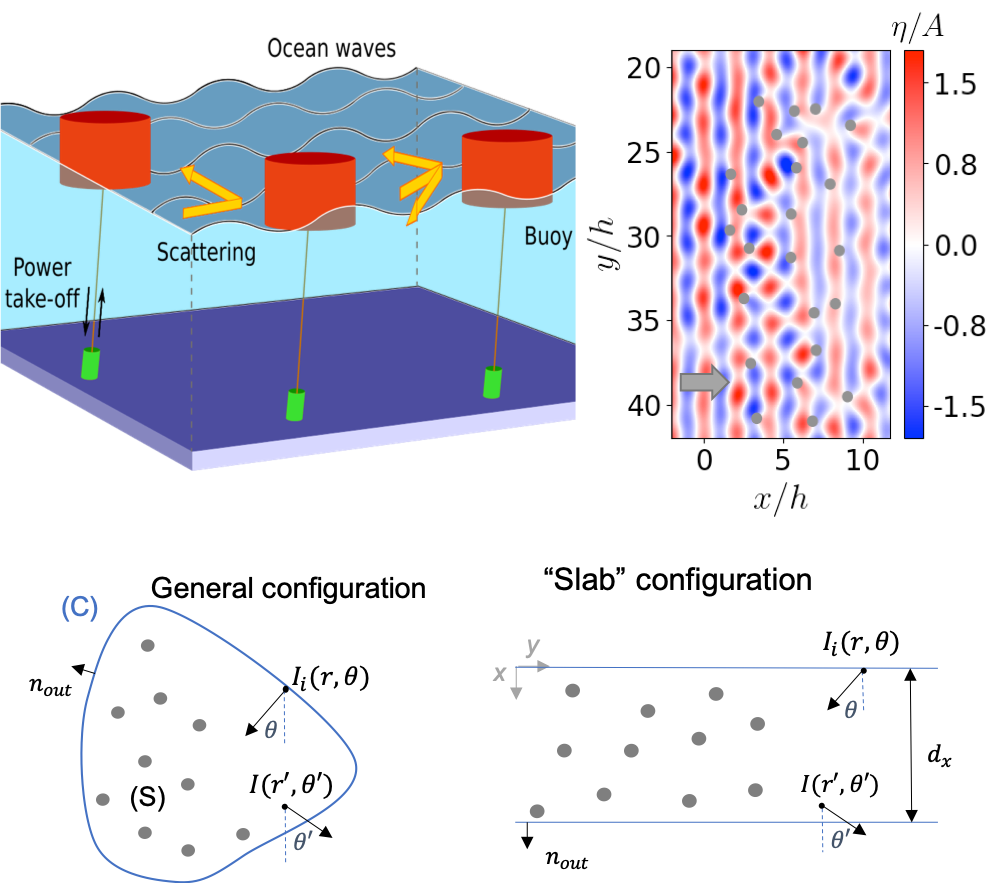}
\caption{Upper left: We bound absorption for very general arrays of ``particles'', including arrays of buoys that extract energy from ocean waves. Upper right: Ocean surface displacement $\eta$ for a cylindrical buoy array \cite{grgur2016optimal} where $A$ is the amplitude of waves incident from left (arrow). Lower: Sketch of RTE system.}
\end{figure}

In optics contexts, scattering particles can be used to enhance absorption in thin-film or dye-sensitized solar cells \cite{nagel2010enhanced,wang2010enhancing,rothenberger1999contribution,galvez2014dye}. Most previous work focused on numerical optimization using the full-wave equations \cite{nagel2010enhanced,wang2010enhancing} or, in the case of dye-sensitized solar cells, RTE for random arrays \cite{rothenberger1999contribution,galvez2014dye}. In \citeasnoun{mupparapu2015path}, approximate analytical estimations of absorption enhancement were given in cases of optically-thin/thick layers under assumptions of weak absorption, normal incidence and isotropic differential cross section. In this work, we were actually motivated by arrays of buoys designed to extract energy from \emph{ocean} waves \cite{falnes2007review,tollefson2014blue, stratigaki2014wave, penesis2016performance} depicted in Fig.~1. Previous numerical-optimization work \cite{cruz2009wave, child2010optimal,  grgur2016optimal, tokic2019hydrodynamics}, in particular a recent extensive computational study on large arrays \cite{grgur2016optimal, tokic2019hydrodynamics}, showed promising results through the design of buoy positions. The question we are trying to answer in this work is more general:  given the absorbing/scattering properties of an individual metaparticle, is there a limit on the total enhancement and how can it be reached? The Yablonovitch limit cannot be applied to all metaparticle arrays since it requires an effective-medium approximation, which is only accurate for either dilute weakly interacting dipolar particles \cite{choy2015effective} or for strongly interacting particles with sufficiently subwavelength separation \cite{smith2002determination}, neither of which is true of the ocean-power problem. Moreover, the Yablonovitch limit is independent of the precise nature of the scattering texture, whereas in our case the whole point is to extrapolate the array properties from the individual-scatterer properties. 
\\

In this paper, we define the \emph{interaction factor} $q(\theta)$ \cite{falnes1980radiation, evans1981power} as the ratio of the power extracted by the array to that of the equivalent number of isolated particles for a given incident angle $\theta$. We first point out that previously known limits in both solar cells and ocean buoys arise from reciprocity constraints on the full-wave equations (Section II). The use of reciprocity in the radiative-transfer equation leads to a general limit (Eq.~\ref{eq6}), valid for any geometrical configuration in RTE regime, that is reached through an isotropic distribution of intensity in the ideal case of small absorption (Section III). This optimal solution justifies the use of a \emph{corrected} radiative-diffusion model (Eq.~\ref{eq1}) that predicts the frequency-averaged performance of random arrays, but also the angle/frequency-averaged performance of the optimized \emph{periodic} array with better than 5\% accuracy. This corrected model can be used to estimate the upper bound on $q$ (which is proportional to the spatially-averaged intensity in RTE framework) even in regimes where the standard diffusion model is not expected to be accurate. This result allows us to quickly evaluate the performance benefits of different metaparticle designs and array configurations, and we show that substantial improvement is possible if the scattering cross-section is increased (relative to the absorption cross-section) and/or if partially reflecting strips are placed on either side of the array (Section IV). More specifically, we show that the use of bending membranes on the water's surface around the buoys significantly increases the interaction factor. We finally use the corrected radiative-diffusion model to find optimal parameters that maximize $q$. 

\section{Reciprocity} The original intuition behind the ray-optical Yablonovitch limit is that the optimal enhancement is achieved through an \emph{isotropic} distribution of light inside the device \cite{yablonovitch1982statistical,green2002lambertian}. This can be thought of as a reciprocity condition. Reciprocity \cite{tsang2000scattering} implies that rays at a given position cannot emerge in the same direction from two different paths. In consequence, if a given point in the absorber is to be reached from as many ray bounces as possible, the rays must be entering/exiting that point from all angles. More formally, we show in \SII{A} that reciprocity can be applied to the \emph{full} Maxwell's equations in order to relate the enhancement to the density of states (accomplished in another way by \citeasnoun{buddhiraju2017theory}), leading to:
\begin{equation} \langle q \rangle = \int_{4\pi}q(\theta)f(\theta)d\Omega \leq \frac{4\pi}{n} \frac{\langle \rho_d \rangle}{\rho_v} \max_{\theta}f   \end{equation}
where $\langle q \rangle$ refers here to the absorption enhancement compared to the single pass, averaged over both polarizations and over a directional spectrum $f(\theta)$ with normalized flux ($\int_{4\pi}|\cos \theta|f(\theta)d\Omega = 1$), $\langle \rho_d \rangle$ is the average density of states in the device, $\rho_v$ the free space density of states and $n$ the index of the absorbing medium. The previous equation becomes an equality in the case of isotropic incidence and small absorption. Yablonovitch limit can then be recovered in bulk media ($\rho_d = n^3 \rho_v$) for an incident field confined to a cone of aperture $2\theta_i$ ($f = \frac{1}{\pi \sin \theta_i^2}\delta(\theta<\theta_i)$): $\langle q \rangle \leq \frac{4n^2}{\sin \theta_i^2}$.

A similar procedure can be followed in the ocean-buoy problem. By applying the appropriate reciprocity relation derived from the wave equation, the Haskind--Hanaoka formula \cite{chiang2005theory}, to the absorption of an \emph{optimal} array of buoys \cite{falnes1980radiation}, one can bound the interaction factor $\langle q \rangle$ for a given directional spectrum $f(\theta)$ ($\int_{2\pi}f(\theta)d\theta$=1) by \citeSII{wolgamot2012interaction}{B}:
\begin{equation} \langle q\rangle = \int_{2\pi}q(\theta)f(\theta)d\theta \leq \frac{M}{k\sigma_a }2\pi \max_\theta f \end{equation}
where $k$ is the wavenumber, $\sigma_a$ the single-buoy absorption and $M$ the number of degrees of freedom for the buoy motion (1--6, e.g. 1 for only heave motion). This result implies that for \emph{isotropic incidence}, we have $\langle q \rangle \leq 1$ at the resonance frequency (the frequency at which the single buoy reaches its maximum absorption $M/k$), while it can in principle be larger at other frequencies. Although this sets a general limit valid at any frequency for any structure, we show in the following that it is not tight when considering the frequency-averaged performance. 

\section{RTE limits} 

We consider a two-dimensional array of scattering/absorbing particles distributed inside a \emph{region} $S$ bounded with a curve $C$ (Fig.~1). 

In the case of dilute and non-structured arrays, coherent scattering effects average out. This allows one to use the radiative-transfer equation (RTE) that only involves specific intensity $I(\vec{r},\theta)$, and that is applicable to ensemble averages of random arrays at a single frequency \cite{ishimaru1978wave, tsang2000scattering}: \begin{equation}\label{eqrte} \vec{e}_\theta \cdot \nabla_rI = -\rho\sigma_eI + \rho\sigma_s \int d\theta'p(\theta,\theta')I+\epsilon \end{equation} where $\sigma_s$, $\sigma_a$ and $\sigma_e$ denote respectively the scattering, absorbtion and extinction cross sections of the individual particles ($\sigma_e=\sigma_s+\sigma_a$), $p$ the normalized differential cross section, $\rho$ the particles' density, $\vec{e}_\theta$ the unit vector with direction $\theta$ and $\epsilon$ internal sources.
  
We conjecture that a similar averaging of coherent effects arises from averaging over frequency and/or angle, and below we demonstrate numerically that this allows RTE to make accurate predictions even for a small number of random samples or for optimized periodic arrays. This is similar to optical light trapping where Yablonovitch model can predict the frequency/angle-average performance of textured solar cells even though it cannot reproduce the exact spectral or angular response \cite{yu2010fundamentalopt, sheng2011optimization}.

\subsection{General limit} 

Similarly to our previous discussion of reciprocity-based limits from the wave equation, we now use reciprocity constraints on RTE to obtain general limits on the interaction factor $q$. 

One can first define a surface Green's function $G_s(\vec{r},\theta;\vec{r'},\theta')$ \cite{case1967linear} giving $I(\vec{r},\theta)$ for an incident field $I_i(\vec{r_i},\theta_i)=\delta (\theta_i-\theta') \delta (\vec{r_i}-\vec{r'})$ and no internal sources $\epsilon = 0$. Similarly, a volume Green's function $G_p(\vec{r},\theta;\vec{r'},\theta')$ can be defined as the intensity $I(\vec{r},\theta)$ obtained with no incident field $I_i=0$ and a point source $\epsilon(\vec{r_i},\theta_i) = \delta (\theta_i-\theta') \delta (\vec{r_i}-\vec{r'})$.

We recall that the flux density $\vec{F}$ is defined as $\int_{2\pi}I\vec{e}_\theta d\theta$. Conservation of energy \cite{ishimaru1978wave} then leads to $\int_C \vec{F}\cdot \vec{n}_{out}d\vec{r}=P_e-P_a$ where $P_e$ and $P_a$ are the generated and absorbed power respectively. For a unit source, we have $P_e=\int_S\epsilon(\vec{r},\theta) d\vec{r}d\theta=1$ so that: \begin{equation}\label{p-bound} \int_C\int_{\vec{e_{\theta}}\cdot \vec{n_{out}>0}}G_p(\vec{r},\theta;\vec{r'},\theta')(\vec{e}_\theta\cdot \vec{n}_{out})d\vec{r}d\theta = 1-P_a \end{equation}

To bound this last expression, we need a \textit{lower}-bound for $P_a$. By noting that the intensity at any point is larger than the \emph{single pass} value (obtained after extinction without multiple scattering), we have: \begin{equation} \label{abs-bound} \begin{split} P_a &= \rho \sigma_a \int_S\int_{2\pi}G_p(\vec{r},\theta;\vec{r'},\theta')d\vec{r}d\theta \\& \geq \rho \sigma_a\int_S \frac{e^{-\rho \sigma_e|\vec{r}-\vec{r'}|}}{|\vec{r}-\vec{r'}|}\delta[angle(\vec{r}-\vec{r'})-\theta']d\vec{r}\\& = \frac{\sigma_a}{\sigma_e} H_{\rho\sigma_e}(\vec{r'},\theta') \end{split} \end{equation} 
where $H_{\rho \sigma_e}(\vec{r'},\theta')$ defined in the previous equation can be interpreted as the power absorbed by a medium \emph{without} scattering and with an absorption coefficient $\rho\sigma_e$ in the presence of a unit source at the point $\vec{r'}$ emitting in direction $\theta'$.

Finally, we relate $G_s$ to $G_p$ through reciprocity using $G_p(\vec{r},\theta;\vec{r'},\theta')|\vec{e}_\theta\cdot \vec{n}_{out}| = G_s(\vec{r'},\pi-\theta';\vec{r},\pi-\theta)$ \cite{case1967linear}. We conclude from Eq. (\ref{p-bound}) and Eq. (\ref{abs-bound}) after a simple change of variable that: \begin{equation} \int_C\int_{\vec{e_{\theta}}\cdot \vec{n_{out}<0}}G_s(\vec{r'},\theta';\vec{r},\theta)d\vec{r}d\theta \leq 1-\frac{\sigma_a}{\sigma_e} H_{\rho\sigma_e}(\vec{r'},\pi-\theta') \end{equation} 
with equality \emph{always} realized in the absence of absorption.

\begin{figure}
\includegraphics[width=0.5\textwidth]{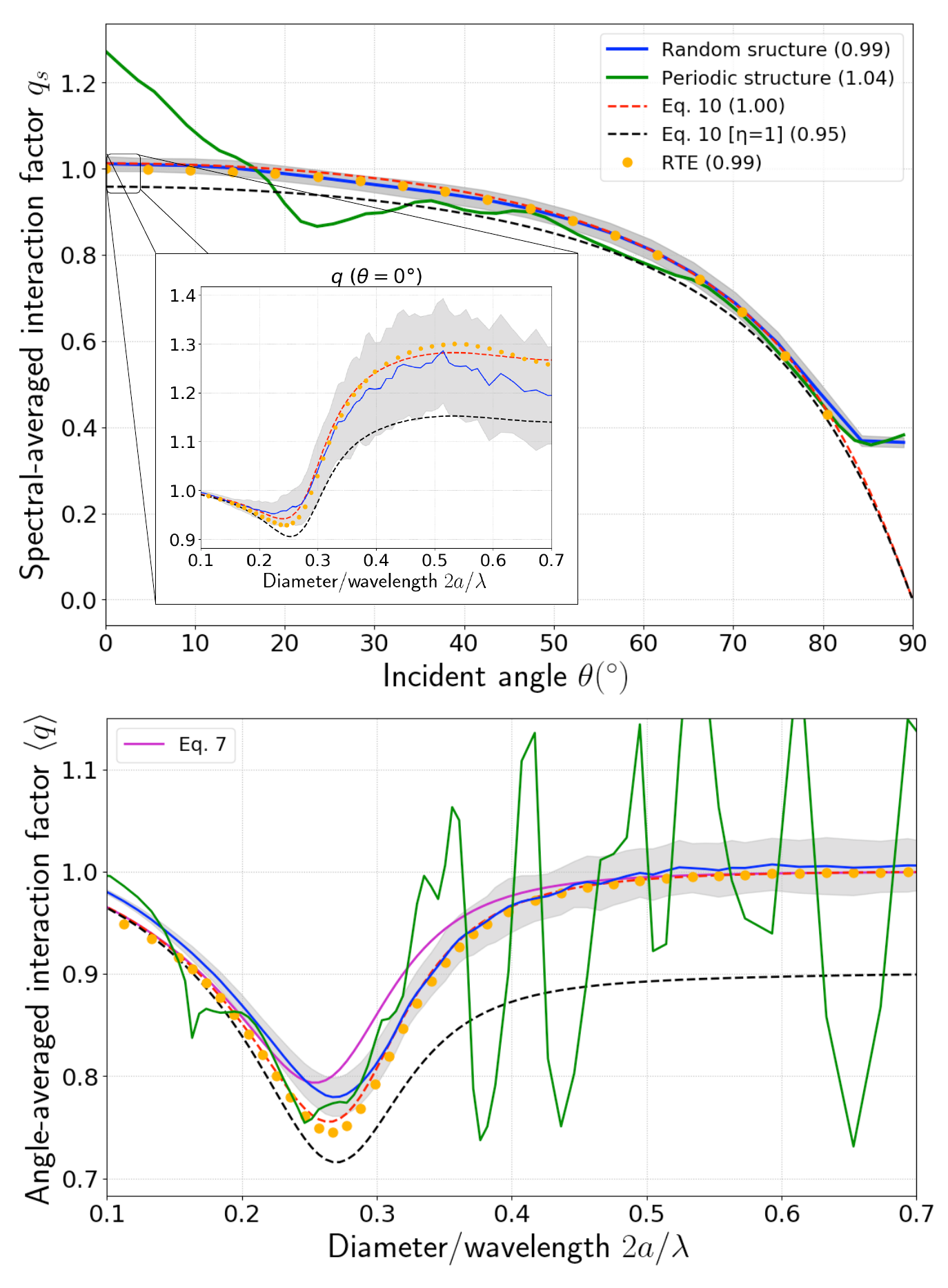}
\caption{Upper: Frequency-averaged interaction factor $q_s$ vs incident angle $\theta$ for $N_x \times N_y = 3 \times 30$ arrays of buoys from exact solution \cite{grgur2016optimal} (solid lines), compared to standard-diffusion (black dashed lines), corrected-diffusion (red dashed lines) and radiative-transfer (RTE with Monte Carlo simulation, dots) models. ($q$ = array absorption / isolated-buoys absorption.)  The average buoy spacings (randomly chosen via a Gamma distribution) are $d_x/h = 1.73$, $d_y/h = 3.63$, with $h$ = ocean depth (the density is $\rho=1/d_yd_x$). Numbers in legend are $q_s$ averaged over $\theta$ for a typical ocean-wave directional spectrum $\cos^{2s}\theta$ with $s=4$ \cite{mitsuyasu1975observations}.  Inset: $q$ vs. wavelength at $\theta=0$, where shaded regions is one standard deviation from mean value (blue line) for 100 random structures. Lower: $\langle q \rangle$ for over isotropic incidence. Results compared to limit in Eq.~(\ref{eq6}).}
\end{figure}

Since the interaction factor in RTE is given by $q = \langle \int_0^{2\pi} I(\vec{r'},\theta')d\theta'\rangle_{\vec{r'}}/I_i$ where $I_i$ is the incident intensity and $\langle . \rangle_\vec{r'}$ is the average over $\vec{r'}$ in S, we can therefore bound the interaction factor $q$ for a given directional spectrum $f(\theta)$ [fraction of power incident from angle $\theta$]:
\begin{equation} \label{eq6} \begin{split} \langle q \rangle & = \int_{C}\int_{2\pi}\int_{\vec{e_{\theta}}\cdot \vec{n_{out}<0}} f(\theta) \langle G_s(\vec{r'},\theta';\vec{r},\theta)\rangle_{\vec{r'}} d\vec{r}d\theta' d\theta \\& \leq 2\pi \left[ 1- \frac{\sigma_a}{\sigma_e} h(\rho\sigma_e)\right] \max_{\theta} f \end{split} \end{equation} 
where $h(\rho\sigma_e) = \langle H_{\rho\sigma_e}(\vec{r'},\theta')\rangle_{\vec{r'},\theta'}\geq 0$. In the case of a ``slab" of thickness $d$, we can show that [\SII{C}]: \begin{equation} h(x) = 1-\frac 2 \pi \frac{1-e_1(xd)}{xd}, e_i(x)=\int_0^{\pi/2}e^{-x\sec \alpha}\cos^i\alpha d\alpha\end{equation}

Note that the bound in Eq.~(\ref{eq6}) reaches its maximal value $2\pi \max f$ in the limit of small absorption. This maximal value, which does not assume optimal single-buoy absorption, generalizes then the previous ocean-buoy bound, giving $\langle q\rangle \leq 1$ for isotropic incidence $f = 1/2\pi$ at any wavelength in RTE regime. In addition, $\langle q\rangle = 1$ is always realized in the small absorption limit. This special case is sometimes referred to as Aronson's theorem \cite{aronson1971theorem}. 

The equality in Eq.~(\ref{eq6}) is reached for: \begin{equation}\int\limits_{C_{\{\vec{e_{\theta}}\cdot \vec{n_{out}}<0\}}}\hspace{-1.5em}G_s(\vec{r'},\theta';\vec{r},\theta)d\vec{r} = \left[1-\frac{\sigma_a}{\sigma_e}H_{\rho\sigma_e}(\vec{r}',\theta')\right]\delta(\theta-\theta_m)\end{equation} where $\theta_m = \argmax f$. This means that the interaction factor should be equal to zero for any incident angle different from $\theta_m$. In the ideal case of small absorption, the optimal $G_s$ becomes independent of $\theta'$, which corresponds to \emph{isotropic} interior intensity, similar to the Yablonovitch model. Therefore, in order to explore \emph{optimal} solutions of RTE, we solve it under the assumption of nearly isotropic intensity, which is well known to lead to a diffusion model \cite{case1967linear, ishimaru1978wave,tsang2000scattering}. We emphasize that not all RTE systems are diffusive, but our result above shows that the optimal $\langle q \rangle$ is attained in an isotropic diffusive regime.

\subsection{Radiative-diffusion model}

Unless otherwise stated, we restrict ourselves to scatterers distributed inside a slab of thickness $d$ (Fig.~1). 

In addition to RTE parameters and reflection coefficients at the boundaries ($R_i$), the radiative-diffusion solution uses an asymmetry factor ($\mu$) \citeSII{joseph1976delta}{F} of the single particle (Fig.~3). The intensity is then given by $ I = I_{ri} + I_d $: $I_{ri}$ is the reduced intensity, solution of $\cos\theta \partial_xI_{ri} = -\rho\sigma_eI_{ri}$, and $I_d$ is the diffuse intensity approximated by $U(x)+\frac 1 \pi \vec{F}(x)\cdot \vec{e}_\theta$ where $U$ verifies a diffusion equation with flux-matching boundary conditions [\SII{D}]. By defining the cross sections per unit of length as $\upsilon_{s,a,e} = \rho d\sigma_{s,a,e}$, the model predicts an interaction factor $q$ of:  \begin{equation}\label{eq1} q (\theta)= q_0(\theta)\left(\eta \left[ D\frac{\xi(\upsilon_d)}{\xi(\upsilon_e\sec\theta)}+C\right]+1\right)\end{equation}  where $\upsilon_d^2=\gamma \upsilon_a(\upsilon_e-\upsilon_s\mu)$ is the diffusion coefficient [$\gamma = 2$ (resp. $=3$) in 2D (resp. 3D)], $\xi(x)$ is the function $(1-e^{-x})/x$, $C =\gamma [\upsilon_s(\upsilon_e+\mu\upsilon_a)]/[\upsilon_d^2-(\upsilon_e\sec \theta)^2]$, $D$ is given by the boundary conditions,  $q_0(\theta)$ is the reduced factor and $\eta$ is an additional correction term that we discuss later. General formulas for $q_0(\theta)$ and $D$ are given in \SII{E}, but in the absence of reflecting walls ($R_i=0$) they simplify to $q_0(\theta) = \xi(\upsilon_e\sec\theta)$ and: \begin{equation} D = -\frac{C(1+e^{-\upsilon_e\sec\theta})+\beta \frac{ (C+\gamma p_1\cos^2\theta)}{(1-p_1)\cos\theta}(1-e^{-\upsilon_e\sec\theta})}{(1+e^{-\upsilon_d})+\beta \frac{\upsilon_d}{\upsilon_e(1-p_1)}(1-e^{-\upsilon_d})},\end{equation}where $p_1=\sigma_s\mu/\sigma_e$ and $\beta = \pi/4$ (resp. $=1$) in 2D (resp. 3D). 

Equation (\ref{eq1}) with $\eta=1$ is obtained from the standard diffusion model. However, it is also known that the diffusion solution is inaccurate for small thicknesses \cite{kim2011correcting,chen2015extending,tricoli2018optimized}. A major problem is that it does \emph{not} guarantee $\langle q\rangle=1$ for isotropic incidence and negligible absorption, even though we previously mentioned that this is the case for any solution of RTE. The reason behind this problem is that the term $I_{ri}$ is not isotropic even for an isotropic incidence. For large thicknesses, however, the contribution of the term $I_{ri}$ becomes negligible and the diffuse term $I_d$ can ensure an isotropic solution. This simply means that the higher order terms in the expression of $I_d$ cannot be neglected for small thicknesses. In order to keep the simplicity of the diffusion solution, we suppose that the effects of higher order terms can be incorporated by the introduction of a scalar term in the diffuse intensity $\eta I_d$ instead of $I_d$. $\eta$ is then defined so as ensure the condition $\langle q\rangle=1$ for isotropic incidence and zero absorption. This procedure is somewhat similar to the approach in \citeasnoun{chen2015extending} except that we use a constant factor $\eta$ since we are interested in the total $q$ and not the spatially resolved $I$. In order to define $\eta$, we study the limit of negligible absorption for which $\upsilon_d \to 0$, $C \to -2\cos^2\theta$ and $D\to \cos^2\theta(1-e^{-\upsilon_e\sec\theta})+\frac \pi 4 \cos\theta (1-e^{-\upsilon_e\sec\theta})$. After simplification, the condition $\langle q\rangle=1$ allows to define $\eta$ as: \begin{equation}\label{eq2} \eta = \frac{\frac \pi 2-\frac{1}{\upsilon_e}[1-e_{1}(\upsilon_e)]}{\beta+\frac \pi 8 \gamma-\frac{2\gamma}{3\upsilon_e}-\beta e_1(\upsilon_e)+\frac \gamma 2 e_2(\upsilon_e)+\frac{\gamma}{\upsilon_e}e_3(\upsilon_e)}.\end{equation}  We note that, as expected, $\eta \to 1$ for an absorber that is thick compared to the extinction length. From our discussion above, this corrected radiative-diffusion model can now be used to estimate the upper bound on the interaction factor even in regimes where the standard diffusion model is not expected to be accurate (optically thin or large absorption).

\section{Ocean-buoy arrays} 

\subsection{Example} We now present a validation of the accuracy of Eq.~(\ref{eq1}) in a model of ocean-wave energy converter (WEC) consisting of a truncated cylinder in heave motion (Fig.~1). The isolated-buoy properties can be obtained analytically \cite{garrett1971wave, yeung1981added, bhatta2003scattering} and are depicted in Fig.~3: they are designed \cite{grgur2016optimal} to have an absorption resonance that matches the typical Bretschneider spectrum \cite{bretschneider1959wave} of ocean waves. We choose the array density based on an earlier optimized periodic 3-row WEC arrangement \cite{grgur2016optimal}.  For this density, we then compare the exact numerical scattering solution calculated for both random and optimized-periodic arrays (using the method from \citeasnoun{grgur2016optimal}) to both the analytical radiation-diffusion $q$ from Eq.~(\ref{eq1}), with and without the correction $\eta$, and the numerical solution of RTE model by a Monte Carlo method \cite{marchuk2013monte}.

In Fig.~2 (upper plot), our corrected model agrees to $< 2\%$ accuracy with exact solutions for random arrays at $\theta < 80^\circ$, as long as the results are frequency-averaged. The importance of frequency averaging is shown by the $q$ frequency spectrum shown in the inset for $\theta=0^\circ$. For an ensemble of random structures, this spectrum exhibits a large standard deviation (gray shaded region), due to the many resonance peaks that are typical of absorption by randomized thin films \cite{yu2010fundamental, sheng2011optimization}, but the \emph{frequency average} mostly eliminates this variance and matches our predicted $q(\theta)$. Precisely such an average over many resonances is what allows the Yablonovitch model to accurately predict the performance of textured solar cells even though it cannot reproduce the detailed spectrum \cite{yu2010fundamentalopt, sheng2011optimization}. 

At first glance, our model does \emph{not} agree in Fig.~2 with the performance of the optimized periodic array from \citeasnoun{grgur2016optimal}: the periodic array, which was optimized for waves near normal incidence, is better at $\theta$ near $0^\circ$ and worse elsewhere. However, when we \emph{also} average over $\theta$ (from a typical ocean-wave directional spectrum \cite{mitsuyasu1975observations}), the result (shown as a parenthesized number in the legend of Fig.~2) matches Eq.~(\ref{eq1}) within 5\%.  If we average over all angles assuming an \emph{isotropic} distribution of incident waves, the results match within 1\%.  Similar results have been observed for thin-film solar cells, in which an optimized structure can easily exceed the $4n^2$ Yablonovitch limit for particular incident angles, but the Yablonovitch result is recovered upon angle-averaging \cite{yu2010fundamentalopt, yu2011angular, sheng2011optimization, ganapati2014light}.

Finally, we note in Fig.~2 (lower plot) that RTE results respect indeed the bound in Eq.~(\ref{eq6}) for isotropic incidence. In particular, we confirm that random arrays achieve $\langle q \rangle = 1$ for small absorption (i.e. small wavelength in our case). The periodic array, on the other hand, doesn't satisfy this relation unless it is frequency averaged. We also mention that the limit Eq.~(\ref{eq6}) is very loose for anisotropic incidence and cannot be reached without using external reflectors as discussed in Section IV-B below.

\subsection{Larger interaction factor} 

\begin{figure}
\includegraphics[width=0.5\textwidth]{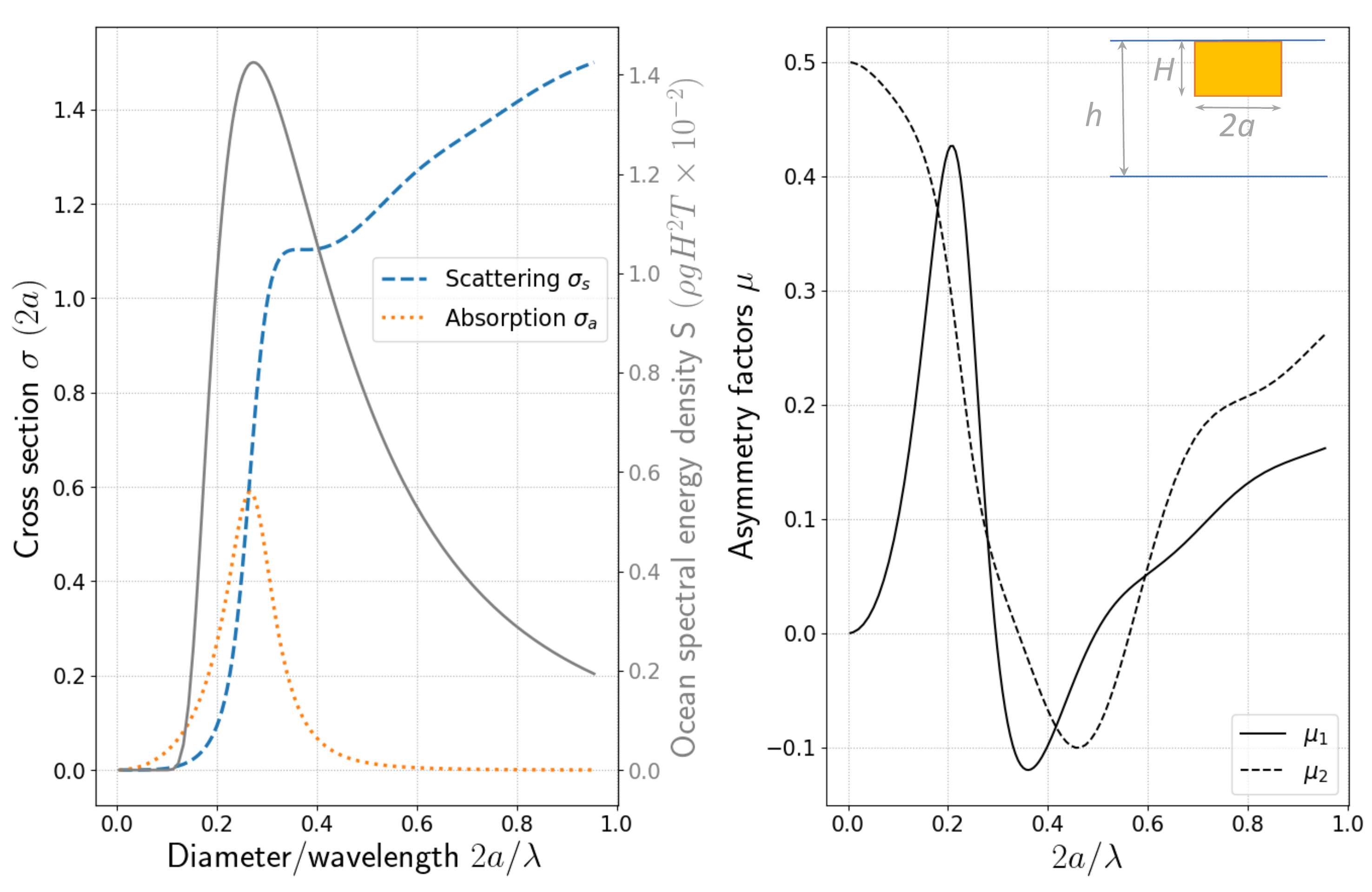}
\caption{Properties of a single truncated-cylinder wave energy converter (WEC) in heave (vertical) motion, with radius $a=0.3h$ and draft $H=0.2h$ where $h$ is the ocean depth. The WEC has an isotropic response with respect to the direction of the incident field. Left: Scattering and absorption cross sections of a single buoy normalized to the cylinder diameter ($\sigma/2a$). The ocean spectral energy density (energy per horizontal surface) is chosen as Bretschneider \cite{bretschneider1959wave} with resonant frequency matching that of the body and is shown in units of $\rho g H^2T$ ($\rho$ is the water density, $g$ the acceleration of gravity, $T$ the mean wave period and $H$ the significant wave height). Right: Asymmetry factors, defined as the average of $\cos\phi$ and $\cos2\phi$ for the two-dimensional differential scattering cross section. These parameters enter into the diffusion equation as $\mu  = (\mu_1-\mu_2)/(1-\mu_2)$ and with $\sigma_s$ replaced by $\sigma_s (1-\mu_2)$ \citeSII{joseph1976delta}{F}.}
\end{figure}

Given this model, we can now explore ways to increase the interaction factor $q$. By examining the dependence of $q$ in Eq.~(\ref{eq1}) on the parameters (Fig.~4), we find that for a fixed scattering-to-absorption ratio $\sigma_s/\sigma_a$, $q$ reaches a maximum $q_{max}$ for an intermediate value of scattering per unit length $ \rho d\sigma_s$, whereas it increases monotonically with $\mu$. A maximum $\mu$ is achieved by increasing $\mu_1$ (forward scattering) and decreasing $\mu_2$ (lateral scattering). The optimal value of $\rho d\sigma_s $ and $q_{max}$ both increase with $\sigma_s/\sigma_a$; as the single particle absorbs more, the interaction factor decreases and the optimal configuration requires a larger spacing between the particles. The maximum $q$ is then achieved in the limit of small absorption ($ \rho d\sigma_a\ll 1$) and large scattering ($ \rho d\sigma_s\gg 1$) for which we obtain a perfect isotropic diffuse intensity. 

From Fig.~3, we see that we have $\sigma_a/\sigma_s \approx 1$ at the resonance of the WEC. In this case, the enhancement is expected to be smaller than 1 around the resonance and optimal structures will tend to have a large spacing $d_y$. (If the array were optimized for small wavelengths $\lambda$, where $\sigma_s \gg \sigma_a$, then a larger $q$ could be obtained at those wavelengths, but $q_s$ would be worse because the optimal spacing in this case is too small for good performance at the resonance.) Still, multiple scattering significantly improves the broadband performance of our array: our $\langle q \rangle \approx 0.99$ is larger than the $\langle q_0 \rangle \approx 0.78$ that is obtained from RTE in absence of multiple scattering (reduced factor $q_0$). The performance is still lower than the 1.65 that would be obtained for $\sigma_s \gg \sigma_a$ in the ideal isotropic regime discussed below, essentially because $\sigma_a/\sigma_s$ is too small and the structure is too thin (as for example quantified by the transport mean-free path $d/l_{tr}={\upsilon_s(1-\mu)} \approx 0.5$ for $\frac{2a}{\lambda} \gtrsim 0.3$) to practically achieve an isotropic diffuse intensity.

Alternatively, we show that $q$ can be enhanced by putting partially reflecting strips around the array. Similar to light-trapping by total internal reflection \cite{yablonovitch1982statistical,green2002lambertian}, one possibility is to use a strip of a lower-``index'' \cite{chiang2005theory} medium (compared to the array's ambient medium) on either side of the array. In the ocean-buoy problem, this can for example be achieved by either a depth change or the use of a tension/bending surface membrane which can lead to near-zero index \cite{zhang2014broadband, bobinski2015experimental}. This modifies equations (2--4) with additional reflection coefficients $R_i$, as given in \SII{E}.

In Fig.~4, we show the effect of an increase in the scattering cross section and/or the index contrast for the same array studied before. By combining both effects, a large ($> 3$) spectral interaction factor can be achieved at normal incidence. At the same time, waves incident at large angles will be reflected out, so that the interaction factor integrated over isotropic incidence is still smaller than 1. For a given directional spectrum and scattering cross section of a single buoy, the optimal interaction factor is achieved for a specific value of the index contrast as can be seen in Fig.~4 (right). 

\begin{figure}[t!]
\includegraphics[width=0.5\textwidth]{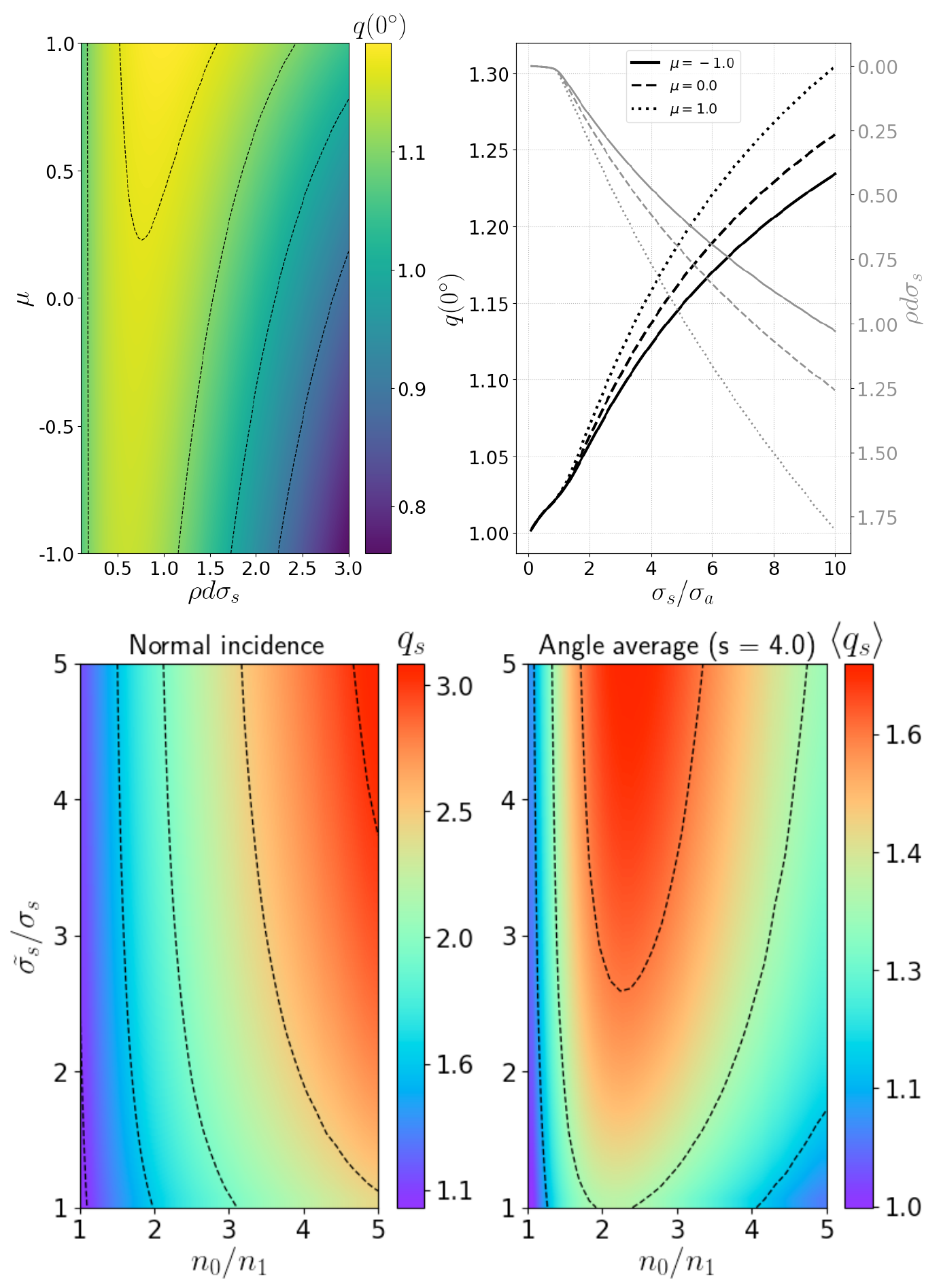}
\caption{Upper: Dependence of $q(0^\circ)$ on parameters in absence of reflecting boundaries. In the left plot, we take $\sigma_s/\sigma_a = 5$. In the right plot, we show the optimal $\rho d\sigma_s$ and $q_{max}$ for different values of $\sigma_s/\sigma_a$ and $\mu$. Lower: Effect of a change in the index contrast and scattering cross section on the bandwidth-averaged factor $q_s$ for the same array in Fig.~2. We tune the index $n_1$ along a strip surrounding the array, with $n_0$ being the index of the array's ambient medium. We suppose that the WEC has new scattering cross section $\tilde{\sigma}_s$, but keep the same absorption cross section. Left: $q_s$ at normal incidence. Right: $q_s$ averaged over $\theta$ with a directional spectrum of $\cos^{2s}\theta$ and s = 4.}
\end{figure}

Finally, it is instructive to look at the ideal case of small absorption and large scattering, for which Eq.~(\ref{eq1}) simplifies to: \begin{equation} \label{eq7}q(\theta) =  \left[1-R_1(\theta)\right] \left(\frac{\pi}{4\alpha}+\cos \theta\right)\cos\theta \end{equation} where $R_1$ is the reflection coefficient of the front-surface and $\alpha = (1-r_1)/(1+r_2)$ with $r_i = \int_{-\pi/2}^{\pi/2}R_1(\theta)\cos^i(\theta)\text{d}\theta / \int_{-\pi/2}^{\pi/2}\cos^i(\theta)\text{d}\theta$. Equation (\ref{eq7}) still gives 1 when averaged over isotropic incidence, but the interaction factor is larger near normal incidence. Without reflectors ($R_1=0$), the maximum value of $q$ at normal incidence is $1+\frac \pi 4$, and the previous directional spectrum gives $\langle q \rangle \approx 1.65$. This maximum value of $q(0)$ does not reach the arbitrarily large enhancement allowed by Eq.~(\ref{eq6}). However, $q(0)$ can still be made sufficiently large by including a reflector designed for transmission near normal incidence and reflection elsewhere (since $\alpha \rightarrow 0$).

\subsection{Surface membrane}

We now use a specific example to demonstrate a larger interaction factor $q$ using surface membranes surrounding the WEC array. For large scale applications, such membranes could be designed to have the desired properties by connecting floating pontoons with elastic elements of appropriate stiffness.

A thin bending membrane on the water surface changes the ``refractive index'' ($\sim k/\omega$) through the following dispersion relation (e.g. \citeasnoun{fox1994oblique}):
 \begin{equation}\label{eq-dispersion} \omega^2 = gk\tanh (kh) \frac{1+C_b(kh)^4}{1+m\cdot kh\tanh(kh)} \end{equation} where $\omega$ is the frequency, $g$ the acceleration of gravity, $k$ the wavenumber, $C_b$ is a dimensionless bending coefficient, $m$ is the mass of the membrane relative to the mass of the water beneath it and $h$ is the water depth. We simply assume $m=0$ in the following. 

At a fixed $\omega$, the membrane decreases $k$ (decreases the ``index'') compared to the surrounding medium. This change of index leads to a reflection off the membrane's edges. In particular, total internal reflection traps the water waves similarly to light trapping in solar cells, which increases the interaction factor $q$. The reflection coefficient, which depends on $\omega$, $C_b$, the incident angle and the membrane's width $w$, can be computed by applying appropriate boundary conditions on either side of the membrane and using a transfer-matrix method as reviewed in \SII{G}. We note that evanescent modes need to be included because of the change in dispersion relations.

The index contrast increases with $C_b$ (increasing stiffness), which increases the range of angles undergoing total internal reflection, making a more effective mirror. Since no waves are coming from the rear of the array, the optimal membrane behind the array should be a perfect reflector ($C_b \to \infty$, limited only by the attainable practical $C_b$). 

\begin{figure}[t]
\includegraphics[width=0.5\textwidth]{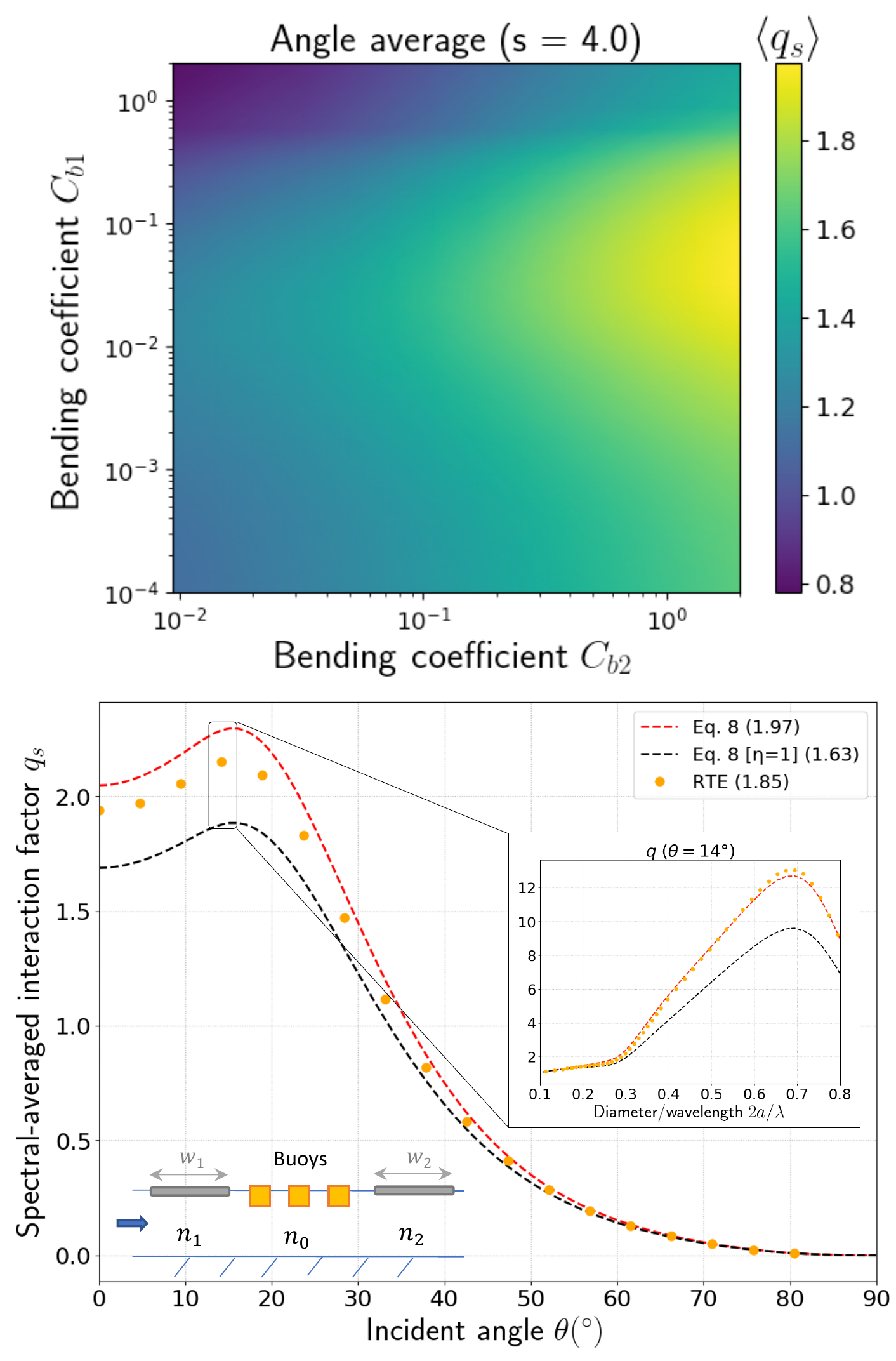}
\caption{Upper: $\langle q_s \rangle$ with a directional spectrum of $\cos^{2s}\theta$ and s = 4 for different values of $C_{b1}$ and $C_{b2}$ corresponding to the front and back membranes respectively. Each point is obtained after optimizing over the membranes' thicknesses. Lower: Frequency-averaged interaction factor $q_s$ vs incident angle $\theta$ for the previously studied array using additional membranes with parameters $(C_{b1},C_{b2})=(0.048,2)$ and $w_1=w_2=1.6h$.} 
\end{figure}

We can now use our corrected diffusion model to predict the upper-bound for the previously studied array as we change $C_{b}$. For each value of $C_{b1}$ and $C_{b2}$ representing the front and rear membranes, respectively, we find the optimal membrane widths that maximize the radiative-diffusion bound. The resulting optimized $\langle q_s\rangle$ values are shown are shown in Fig.~5 (upper plot). We first note that the frequency/angle-averaged interaction factor $\langle q_s\rangle$ increases significantly ($>1.8$) compared to the $\langle q_s\rangle = 1.00$ without the membranes. We also confirm that $\langle q_s\rangle$ increases with $C_{b2}$ (rear membrane) as expected. On the other hand, there is an optimal value for $C_{b1}$ depending on the directional spectrum $f(\theta)$. For a \emph{focused} incident field, only angles near normal incidence matter so that $C_{b1}$ can be increased allowing more of the waves scattered by the WECs to be trapped. On the other hand, for a broad directional spectrum, a large value of $C_{b1}$ prevents waves incident from wide angles from reaching the WECs. 

For our array, supposing for example that the maximal attainable value of $C_{b2}$ is equal to $2$, the optimal value for $C_{b1}$ is $0.048$ with optimal widths equal to $1.6h$ for both the front and rear membranes. The frequency-averaged interaction factor $q_s$ for the optimal parameters is shown in Fig.~5 (lower plot). Our predicted bound (red dashed line = corrected diffusion) is indeed larger than the actual performance of the array as modeled by RTE (orange dots). That is mainly due to the relatively small scattering cross section compared to the absorption cross section. As illustrated in the inset of Fig.~5 at small wavelengths where $\sigma_s$ is large (Fig.~3), we see that an increase in the scattering cross section leads to arrays with performance closer to the radiative-diffusion bound. 

We finally mention that in the case of using a perfect back-reflector, $\langle q_s\rangle$ can reach a value of 2.26 for $C_{b1}=0.06$ and $w_1=1.65h$.

\section{Conclusion.} 
We believe that the angle/frequency-averaged limits presented in this paper provide guidelines for future designs to achieve a large $q$ factor which may open the path for the realization of large arrays of buoys for efficient ocean energy harvesting. In particular, the use of external reflecting elements such as surface membranes seems a promising approach. The results are also applicable to other problems where multiple scattering effects are used to achieve enhancement, including scattering particles inside an absorbing layer. One can, for example, recover the standard Yablonovitch-$4n^2$ result from our approach in an appropriate limit [\SII{H}], but the real power of our result is that it allows to study the effect of single-metaparticle properties, angle of incidence and reflecting boundaries. 

\begin{acknowledgments}
This work was supported in part by the Army Research Office under Cooperative Agreement Number W911NF-18-2-0048. 
\end{acknowledgments}

\appendix

\section{Enhancement from reciprocity of Maxwell's equations}

Although the end result is not new, we wish to emphasize that the underlying ideas of the Yablonovitch and LDOS limits are closely tied to reciprocity.  This is an alternative to the derivation in \citeasnoun{buddhiraju2017theory}, which differs in that it directly uses the reciprocity (or generalized reciprocity) from Maxwell's equations.  As was also emphasized in \citeasnoun{buddhiraju2017theory}, the result also applies to linear \emph{nonreciprocal} systems, since the density of states of transposed-related materials is the same ($G_{\epsilon}(r,r) = G_{\epsilon^t}^t(r,r)$ \cite{tsang2000scattering}).

Here for simplicity, we consider a reciprocal system in the derivation. We have then: \begin{equation}\label{eq18}  \int_{S_{\infty}}[\mathbf{E_a} \times \mathbf{H_b} - \mathbf{E_b} \times \mathbf{H_a} ]\cdot \mathbf{\hat{k}}\; dS = \int_V [\mathbf{E_a}\cdot \mathbf{J_b}-\mathbf{E_b}\cdot \mathbf{J_a}]dV\end{equation}

If we choose ($\mathbf{J_a}=\frac{1}{j\mu \omega}\mathbf{\hat{e}_s}\delta_\mathbf{r_0}$, $\mathbf{E_a^{inc}} = \mathbf{0}$) and ($\mathbf{J_b}=\mathbf{0}$, $\mathbf{E_b^{inc}} = e^{jk\mathbf{\hat{k}_0}\cdot \mathbf{r}}\mathbf{\hat{e}_b}$), then $\mathbf{E_a}  = \mathbf{\Bar{\Bar{G}}}_E(\mathbf{r_0},\mathbf{r_0})\mathbf{\hat{e}_s}$.

The far field term can be written as $\mathbf{E_a^s} = f_s(\mathbf{\hat{k}})\frac{e^{jkr}}{r}\mathbf{\hat{e}_a}, \; \mathbf{H_a^s}=\frac 1 \eta (\mathbf{\hat{k}} \times \mathbf{E_a^s})\; \text{with}\; \eta = \sqrt{\frac{\mu_0}{\epsilon_0}}$, and similarly for the far-field of the scattered field $``b"$, so that: $ \int_{S_{\infty}}[\mathbf{E_a^{s}} \times \mathbf{H_b^{s}} - \mathbf{E_b^{s}} \times \mathbf{H_a^{s}}] \cdot \mathbf{\hat{k}}\; dS = 0$.

We then expand the integrand of the left term in \ref{eq18} to obtain: \begin{equation}\begin{split} & \int_{S_{\infty}}[\mathbf{E_a^{s}} \times \mathbf{H_b^{inc}} - \mathbf{E_b^{inc}} \times \mathbf{H_a^{s}}] = -\frac 1 \eta  \int f_s(\mathbf{\hat{k}}) e^{jkr(1+\mathbf{\hat{k}} \cdot \mathbf{\hat{k}_0})}\\& [( \mathbf{\hat{e}_a} \cdot \mathbf{\hat{e}_b}) (1-\mathbf{\hat{k}}\cdot \mathbf{\hat{k}_0})+(\mathbf{\hat{e}_a} \cdot \mathbf{\hat{k}_0})(\mathbf{\hat{e}_b} \cdot \mathbf{\hat{k}})] rd\mathbf{\hat{k}} \end{split} \end{equation}

The integral can be evaluated using the method of stationary phase \cite{choi2011method}. The function $g(\theta,\phi) = 1+\mathbf{\hat{k}} \cdot \mathbf{\hat{k_0}} = 1+\cos\theta\cos\theta_0+\sin\theta\sin\theta_0\cos(\phi-\phi_0)$ has two extrema at $\pm \mathbf{\hat{k_0}} $. The integrand is null at the first, so only the second matters. The Hessian matrix at $-\mathbf{\hat{k_0}} $ is given by: $\begin{bmatrix} 1 & 0 \\ 0 & \sin\theta_0^2\end{bmatrix}$. We then conclude that the integral we want to evaluate is equal to: \begin{equation}\begin{split}& -\frac1 \eta j \frac{1}{\sin\theta_0/2} \frac{1}{kr}[2(\mathbf{\hat{e}_a} \cdot \mathbf{\hat{e}_b}))-(\mathbf{\hat{e}_a} \cdot \mathbf{\hat{k}_0})(\mathbf{\hat{e}_b} \cdot \mathbf{\hat{k}_0})]\\& f_s(-\mathbf{\hat{k}_0})r\sin\theta_0  = -\frac j \eta  \frac{4\pi}{k} (\mathbf{\hat{e}_a} \cdot \mathbf{\hat{e}_b})f_s(-\mathbf{\hat{k}_0}) \end{split}\end{equation}
where $\mathbf{\hat{e}_a}$ is evaluated at $-\mathbf{\hat{k}_0}$.

We finally conclude from \ref{eq18} that: \begin{equation}\label{eq21} -\mathbf{\hat{e}_s}\cdot \mathbf{E_b}(\mathbf{r_0}) = 4\pi (\mathbf{\hat{e}_a}\cdot\mathbf{\hat{e}_b})f_s(-\mathbf{\hat{k}_0})\end{equation}
which is the reciprocity relation relating the far field of a point source at $\mathbf{r_0}$ in the direction $-\mathbf{\hat{k}_0}$ to the field at $\mathbf{r_0}$ due to an incoming plane wave from the same direction. 

Now, we use the Poynting theorem to compute the far field of the point source: \begin{equation}\label{eq19}\begin{split}\frac 1 \eta \int |f_s(\mathbf{\hat{k}})|^2d\mathbf{k} & = \int Re[\mathbf{E_a}\times \mathbf{H_a^*}]\cdot \mathbf{\hat{k}}dS \\&\leq -\int Re[\mathbf{J_a^*}\cdot \mathbf{E_a}] \\& = \text{Im}[\mathbf{E_a}(\mathbf{r_0})\cdot \mathbf{\hat{e}_s}]\frac{1}{\omega \mu}\end{split} \end{equation}

At this point we are able to combine \ref{eq21} and \ref{eq19} to find our main result about the enhancement. We consider an incoming angular distribution $f(\mathbf{\hat{k}_0})$ with a normalized flux ($\int_{4\pi}|\cos\theta|f(\mathbf{\hat{k}_0})d\mathbf{\hat{k}_0} = 1$). By integrating over all coming angles and polarizations of the ``b" field, we have:\begin{equation}\label{eq20} \begin{split} 
\int \sum_{\mathbf{\hat{e}_b}} & |\mathbf{E_b}|^2f(\mathbf{\hat{k}_0})d\mathbf{\hat{k}_0}  = \int \sum_{\mathbf{\hat{e}_b},\mathbf{\hat{e}_s}} |\mathbf{E_b}\cdot \mathbf{\hat{e}_s}|^2f(\mathbf{\hat{k}_0})d\mathbf{\hat{k}_0} 
\\& = (4\pi)^2\int \sum_{\mathbf{\hat{e}_b},\mathbf{\hat{e}_s}} |\mathbf{\hat{e}_a}\cdot \mathbf{\hat{e}_b}|^2|f_s(-\mathbf{\hat{k}_0})|^2f(\mathbf{\hat{k}_0})d\mathbf{\hat{k}_0} 
\\ & = (4\pi)^2\int \sum_{\mathbf{\hat{e}_s}} |f_s(-\mathbf{\hat{k}_0})|^2f(\mathbf{\hat{k}_0})d\mathbf{\hat{k}_0} 
\\& \leq (4\pi)^2\frac{\max f}{k} \sum_{\mathbf{\hat{e}_s}}  \text{Im}[\mathbf{E_a}(\mathbf{r_0})\cdot \mathbf{\hat{e}_s}] 
\\ & = (4\pi)^2\frac{\max f}{k} \text{Tr}[ \text{Im} \mathbf{\Bar{\Bar{G}}}_E(\mathbf{r_0},\mathbf{r_0})] 
\\& = (4\pi)^2\frac{\max f}{k}  \frac{\pi c^2}{\omega n^2}\rho_d(\mathbf{r_0}) \end{split}\end{equation} 
which relates rigorously the enhancement and the local density of states. 

We can use this result to compute the absorbed power and deduce the enhancement compared to the single pass for a cell of surface $S$ and effective thickness $d$. We have: \begin{equation} \label{eq22} \begin{split} \langle P_{abs} \rangle &= \frac 1 2 \omega \epsilon'' \epsilon_0 \int_{V} \int \sum_{\mathbf{\hat{e}_b}} |\mathbf{E_b}|^2f(\mathbf{\hat{k}_0})d\mathbf{\hat{k}_0}  \\& \leq \frac 1 2 \epsilon'' \epsilon_0 (4\pi)^2 \frac{\pi c^3}{\omega n^2} \max f \int_V\rho_d \end{split}\end{equation}

The total incident power, taking into account the two polarizations, is given by $ \frac{1}{2\eta} \int f(\mathbf{\hat{k}_0}) |\cos\theta| d\mathbf{\hat{k}_0} \times 2 \times S = \frac{S}{\eta}$, and the normalized single pass absorption is $ \alpha d = \frac{\epsilon''}{n} \frac \omega c d$. The enhancement is then given by: \begin{equation} \langle q \rangle = \frac{\langle P_{abs}\rangle }{P_{inc}\alpha d} \leq \frac{4\pi}{n} \frac{\langle \rho_d\rangle}{\rho_v} \max f\end{equation}
where $\rho_v = \frac{\omega^2}{2\pi^2 c^3}$ is the free space density of states. This inequality becomes an \emph{equality} in the case of negligible absorption and isotropic incidence ($f = \frac{1}{2\pi}$).

For a bulk dielectric, we have $\rho_d = n^3 \rho_v$ so that $\langle q\rangle \leq 2n^2$ for isotropic incident light which is the standard limit in the absence of a back reflector.

\section{Interaction factor from reciprocity in ocean waves}

In this section we review the result in \citeasnoun{wolgamot2012interaction} and emphasize that it is also a consequence of reciprocity, which shows the similarity with the LDOS limit in solar cells.

The problem of ocean wave energy extraction using oscillating bodies is formally equivalent to the problem where there are discrete sources of which the amplitude can in principle be controlled externally (velocity of the body that can be controlled through an external mechanical mechanism). Considering the effect of the incoming wave and interaction between bodies, the total absorption can be written as a quadratic function in terms of the amplitudes of the different sources as in \cite{falnes1980radiation} for example. Maximizing the absorption allows to find the optimal amplitudes as a function of the scattered field and the radiated fields from the sources. This gives \cite{falnes1980radiation}: \begin{equation}P_{max} =\frac1 8 \mathbf{F_e}^*(\theta)R^{-1} \mathbf{F_e}(\theta) \end{equation}
where $\mathbf{F_e}(\theta) $ is the force applied on the bodies for an incident wave from the direction $\theta$ and $R$ is the \textit{resistance} matrix (radiation damping matrix). 

One would try to see the effect of the reciprocity relations discussed before on the maximum absorption in this context. The exact equivalent of Eq.~(\ref{eq21}) is already known in the ocean waves problem as the Haskind-Hanaoka formula that relates the force applied on a body due to an incident wave to the radiated field when the the body acts as a source \cite{chiang2005theory}. It leads to:  \begin{equation}\label{eq28}F_{e,i}(\theta)=-\frac 4 k \rho_o g A c_g A_i(\theta+\pi) \end{equation} where $A$ is the amplitude of the incident wave, $A_i$ is the far-field amplitude of the radiation mode $i$, $k$ is the wavenumber, $c_g$ is the group velocity, $\rho_o$ is the water density, and $g$ is the gravity of Earth. 

The use of this formula on the maximum absorbed power by an array of oscillating bodies leads to the bound on the power absorbed by the array. For a given incident angular distribution $f(\theta)$ normalized so that $\int_{2\pi}f(\theta)d\theta=1$: \begin{equation}\begin{split} \langle P_{max} \rangle &= \int f(\theta)P_{max}(\theta)\text{d}\theta \\ & \leq \max f\int P_{max}(\theta)\text{d}\theta \\& = \max f \frac 1 8 \sum_{i,j}[R^{-1}]_{i,j}\int_{2\pi}F_{e,i}^*F_{e,j}\text{d}\theta \end{split} \end{equation}

Using \ref{eq28} and the fact that $R_{i,j} = \frac{2}{\pi k}\rho_o g c_g\; \text{Re}(\int_{2\pi}A_i^*A_j)$ \cite{falnes1980radiation}, we conclude that: \begin{equation}\label{eq10}\langle \sigma_{a,max}^N \rangle = \int_{2\pi} \sigma_{a,max}^N(\theta)f(\theta)d\theta \leq \frac{NM}{k} 2\pi \max f \end{equation}
where $\sigma_{a,max}^N=P_{max}/(\frac 1 2 \rho_og|A|^2c_g)$ is the \emph{maximum} absorption cross section of the array, $N$ is the number of buoys and $M$ is the number of degrees of freedom for the buoy motion (1--6 \cite{falnes1980radiation}, e.g. 1 for only heave motion). 

This result is general and does not depend on assumptions on the scatterers. It means that the interaction factor $q = \langle \sigma_{a,max}^N\rangle /N\langle \sigma_{a}^1\rangle$ is bounded by $M/(k\langle \sigma_a^1\rangle )$ for isotropic incidence. For buoys in heave motion which are studied in this paper, we have $M=1$ and $\langle \sigma_a^1\rangle  = \sigma_a^1$ (the absorption cross section of the single buoy does not depend on the incident angle). 

Note that Eq.~(\ref{eq10}) is also valid for a single buoy. Depending on the symmetries of the buoy, the actual absorption may be smaller (for an axisymmetric buoy, we always have $k \sigma_a^1\leq 3$ \cite{chiang2005theory}).

It is important to realize that this bound is equal to $1$ at the resonance frequency [the $k$ where $\langle \sigma_{a}^1 \rangle$ reaches the maximum $M/k$ from (\ref{eq10})], while it can in principle be larger at other frequencies. 

\section{General RTE limit for a ``slab''}

We compute the function $h$ in Eq.~(\ref{eq6}) for a slab of thickness $d$ (with perfectly transmitting boundaries). We assume that the slab is normal to the \emph{x}-axis.

We first write the integral $H$ using polar coordinates $(r,\theta)$: \begin{equation}\begin{split} & H_\alpha(x',\theta')= \int_{-\pi/2}^{\pi/2}\int_{0}^{\frac{x'}{\cos\theta}}\alpha e^{-\alpha r} \delta(\theta-\theta')d\theta dr \\&+ \int_{-\pi/2}^{\pi/2}\int_{0}^{\frac{d-x'}{\cos\theta}}\alpha e^{-\alpha r} \delta(\theta-\theta')d\theta dr \end{split} \end{equation}

After simplification, we have then: \begin{equation}\begin{split} h(\alpha)& = \frac{1}{2\pi d} \int_{0}^{2\pi} \int_{0}^{d} H_\alpha (x',\theta')dx'd\theta' \\ & = 1-\frac{2}{\pi\alpha d} \left[ 1-\int_0^{\pi/2}e^{-\alpha d\sec \theta}\cos\theta d\theta \right] \end{split} \end{equation} 

\section{Diffusion equation}

Here we reproduce the diffusion equation as in \citeasnoun{ishimaru1978wave, tsang2000scattering} but adjusting the numerical coefficients for a two-dimensional medium.  

We first separate the intensity as $I = I_{ri}+I_{d}$ where $I_{ri}$ is the reduced (coherent) intensity and $I=I_{d}$ is the diffuse (incoherent) intensity. The reduced intensity is related to the single scattering and obeys: $ \vec{e}_\theta \cdot \nabla_rI_{ri}=-\rho \sigma_e I_{ri} $. So from RTE equation, the diffuse intensity obeys: \begin{equation} \label{eqA2} \begin{split}\vec{e}_\theta \cdot \nabla_r I_d &= -\rho \sigma_e I_d + \rho \sigma_s \int d\theta'p(\theta,\theta')I_d + J, \\ J &=  \rho \sigma_s \int d\theta'p(\theta,\theta')I_{ri}\end{split} \end{equation}

Now, considering the diffusion approximation, we write: $I_d(\mathbf{r},\theta) = U(\mathbf{r})+\frac{1}{\pi}\mathbf{F}(\mathbf{r})\cdot \vec{e_\theta}$. This could be seen as a first order series in $\theta$. We also note that the diffuse flux is: $\int I_d\vec{e}_\theta\; d\theta  = \mathbf{F} $.

In order to obtain $U$ and $\mathbf{F}$ we apply the operators $\int d\theta$ and $\int \vec{e}_\theta d\theta$ on (\ref{eqA2}). This leads to: \begin{equation}\label{eqA3}\begin{split} \nabla_r \cdot \mathbf{F} &= -2\pi \rho\sigma_a   U + 2\pi \rho \sigma_s U_{ri}\\ U_{ri}(\mathbf{r}) &= \frac 1 {2\pi} \int d\theta\; I_{ri}(\mathbf{r}, \theta)  \\ \nabla_r U &= -\frac 1 \pi \rho \sigma_{tr} \mathbf{F} + \frac 1 \pi \int d\theta\; J \mathbf{\hat{s}}\end{split}  \end{equation}
where $\sigma_{tr}=\sigma_e(1-p_1)$ and $\sigma_e p_1 =  \int d\theta' p(\mathbf{\hat{s}},\mathbf{\hat{s}'}) [\mathbf{\hat{s}} \cdot \mathbf{\hat{s}'}]$, so that $p_1 = \sigma_s \mu /\sigma_e$ where $\mu$ is the average of the cosine of the scattering angle. 

Equations (\ref{eqA3}) allow to solve for $U$ and $\mathbf{F}$. Combining them, we obtain a diffusion equation for $U$: 
\begin{equation} \nabla^2 U - (\rho \sigma_d)^2 U = -2\rho^2 \sigma_{tr}\sigma_s U_{ri} + \frac 1 \pi \nabla \cdot \int d\theta \; J \mathbf{\hat{s}} \end{equation}

Now we need to add appropriate boundary conditions. Supposing that we have a reflection coefficient $R$ on the surface, this should be: $I_d(\mathbf{r},\theta) = R(\theta) I_d(\mathbf{r},\pi-\theta) $ for $\mathbf{\hat{s}}$ directed towards the inside of the medium. However, considering the assumed formula for $I_d$ the condition cannot be satisfied exactly . A common approximate boundary condition is to verify the relation for the fluxes:
\begin{equation} \int_{\mathbf{\hat{s}}\cdot \mathbf{\hat{n}}>0} I_d(\mathbf{\hat{s}}\cdot \mathbf{\hat{n}}) d\theta= \int_{\mathbf{\hat{s}}\cdot \mathbf{\hat{n}}<0} R(\theta)I_d(\mathbf{\hat{s}}\cdot \mathbf{\hat{n}}) d\theta \end{equation} 
where $\mathbf{\hat{n}}$ is the normal to the surface directed inwards.

Using the formula for $I_d$ we obtain: \begin{equation}  2(1-r_1)U+\frac{(1+r_2)}{2}\mathbf{F}\cdot \mathbf{\hat{n}} = 0 \end{equation}
where $r_i = \int_{-\pi/2}^{\pi/2}R(\theta)\cos^i(\theta)d \theta / \int_{-\pi/2}^{\pi/2}\cos^i(\theta)d \theta	$.

\section{General expression for the interaction factor} 

We give the expression for $q$ in the presence of reflecting boundaries with angle-dependent reflection coefficients $R_i$ ($R_1$ refers to the boundary facing the incident wave).

Using the same notation as in Section III, we have: \begin{equation} q_0(\theta) = \frac{(1-\Tilde{R_1})(1+\Tilde{R_2}Y)}{1-\Tilde{R_1}\Tilde{R_2}Y^2} \xi(\upsilon_e\sec\theta) \end{equation}
with $\Tilde{R_i} = R_i(\theta)$ and $Y = e^{-\upsilon_e\sec\theta}$.

$D$ is given through boundary conditions by $D = \frac{A+B}{1+\Tilde{R_2}Y}$, where: \begin{equation}\begin{split}& \begin{bmatrix} \alpha_1+\beta \frac{\upsilon_d}{\upsilon_{tr}} &  (\alpha_1-\beta \frac{\upsilon_d}{\upsilon_{tr}})e^{-\upsilon_d}  \\ (\alpha_2-\beta \frac{\upsilon_d}{ \upsilon_{tr}})e^{-\upsilon_d} & (\alpha_2+\beta \frac{\upsilon_d}{\upsilon_{tr}})\end{bmatrix} \begin{bmatrix} A \\ B\end{bmatrix}  =X=\\& -\begin{bmatrix} C(1+\Tilde{R_2}Y^2)\alpha_1+\beta \frac{\upsilon_e}{\upsilon_{tr}}(\frac{C}{\cos \theta}+\gamma p_1\cos \theta )(1-\Tilde{R_2}Y^2)\\ [C(1+\Tilde{R_2})\alpha_2-\beta \frac{\upsilon_e}{\upsilon_{tr}}(\frac{C}{\cos \theta}+\gamma p_1 \cos \theta)(1-\Tilde{R_2})] Y\end{bmatrix} \end{split}\end{equation}
with $\upsilon_{tr} = \upsilon_e-\upsilon_s\mu$, $\alpha_i = (1-r^i_1)/(1+r^i_2)$ and $r^i_p = \int_{-\pi/2}^{\pi/2}R_i(\theta)\cos^p(\theta)\text{d}\theta / \int_{-\pi/2}^{\pi/2}\cos^p(\theta)\text{d}\theta$. We recall that ($\gamma = 2$, $\beta = \pi/4$) [resp. ($\gamma = 3$, $\beta = 1$)] in 2D [resp. 3D].

The correction term $\eta$, which ensures that the interaction factor for isotropic incidence and zero absorption is 1, is defined as: 
\begin{equation} \label{eq8} \eta = \frac{\pi-\sum\limits_{i=1}^2\displaystyle\int_0^{\pi/2}q_0^{(i)}d\theta}{\sum\limits_{i=1}^2\displaystyle\int_0^{\pi/2}\left[\frac{q_0^{(i)}D_0^{(i)}(\theta,\upsilon_e,\upsilon_{tr})}{\xi(\upsilon_e\sec\theta)}-\gamma \cos^2\theta q_0^{(i)}\right]d\theta}\end{equation}
with: \begin{equation} \label{eq4}(1+\Tilde{R_2}Y)D_0(\theta,\upsilon_e,\upsilon_{tr}) =\frac{(\alpha_2+\frac{2\beta}{\upsilon_{tr}})X_{0,1}+(\alpha_1+\frac{2\beta}{\upsilon_{tr}})X_{0,2}}{\frac{2\beta}{\upsilon_{tr}}(\alpha_1+\alpha_2)+2\alpha_1\alpha_2}  \end{equation}
where: \begin{equation}X_0= \begin{bmatrix}\gamma \cos^2\theta(1+\Tilde{R_2}Y^2)\alpha_1+2\beta \cos\theta(1-\Tilde{R_2}Y^2)\\ [\gamma \cos^2\theta(1+\Tilde{R_2})\alpha_2-2\beta \cos \theta(1-\Tilde{R_2})] Y\end{bmatrix}\end{equation}
Superscripts for $q_0^{(i)}$ and $D_0^{(i)}$ refer to the boundary that is facing the incident wave.

\section{Asymmetry factor}

The asymmetry factor usually used in diffusion models is \cite{ishimaru1978wave, tsang2000scattering} $\mu=\mu_1$, where in general $\mu_i=\int_{2\pi} \cos(i\theta) p(\theta)d\theta$ (where we take $p(\theta,\theta')=p(\theta-\theta')$). Since the diffusion result depends only on $\upsilon_s$, $\upsilon_a$ and $\mu_1$, it can be seen as approximating the differential scattering cross section by: $p(\theta-\theta')=\frac{1}{2\pi}[1+2\mu_1\cos(\theta-\theta')]$.

The Delta-Eddington approximation \cite{joseph1976delta} allows to incorporate the second moment of $p$ by including the forward scattering peak using a ``delta function" term so that: $p(\theta,\theta')=\mu_2\delta(\theta-\theta')+\frac{1-\mu_2}{2\pi}[1+2\mu\cos(\theta-\theta')]$ where $\mu=(\mu_1-\mu_2)/(1-\mu_2)$. This approximation matches the Fourier decomposition of $p$ up to the second term. By incorporating this expression in RTE (Eq.~\ref{eqrte}), one recovers a \emph{second} RTE with $p$ replaced by $\frac{1}{2\pi}[1+2\mu\cos(\theta-\theta')]$ and $\sigma_s$ replaced by $\sigma_s(1-\mu_2)$. So the diffusion approximation can be made more accurate by replacing $\mu$ by $(\mu_1-\mu_2)/(1-\mu_2)$ and $\sigma_s$ by $\sigma_s (1-\mu_2)$. This is known as the Delta-Eddington approximation \cite{joseph1976delta}.

In a three-dimensional medium, $\mu_i=\int_{4\pi} P_i(\cos\theta) p(\cos\theta)d\Omega$ where $P_i$ is the $i^{th}$ Legendre polynomial.

 \section{Reflection coefficient with membranes}
  
 We consider a plane wave arriving from medium (1), that is a free-surface ocean with finite depth $h$, at angle $\theta$ with respect to the $x$-axis. We suppose that we have a thin membrane (2) on the water surface extended from $x=0$ to $x=w$. Change in the dispersion relation leads to different wavenumbers $k_{i}^n$ verifying: \begin{equation} \omega^2 = gk_1^{n}h\tanh (k_1^{n}h) = gk_2^{n}h\tanh (k_2^{n}h)(1+C_b(k_2^{n}h)^4) \end{equation}
 where $C_b$ is a bending coefficient of the membrane. $k_i^0$ corresponds to a (real) propagating wave while the other $k_i^n$ correspond to (pure imaginary) evanescent waves.
 
 We first compute the transfer-matrix between medium (1) and medium (2). We write the velocity potential in each medium $i$ as: \begin{equation} \phi_i = \sum_{n=0}^N f_{n,i}(z) \left[ \alpha_{n,i}e^{ik_{x,i}^n}+\beta_{n,i}e^{-ik_{x,i}^n} \right]e^{ik_yy}\end{equation}
 where $(k_{x,i}^n)^2+k_y^2=(k_i^n)^2$ and $f_{n,i}(z) = N_{n,i} \cosh k_i^n(z+h)$ ($z=0$ is the water's free surface). $N_{n,i} = 1/\sqrt{1+\frac{\sinh(2k_i^nh)}{2k_i^nh}}$ is defined so as to ensure that $\langle f_{n,i}, f_{n,i} \rangle= \int_{-h}^{0}f_{n,i}^2dz=1$. We also note that $(f_{n,1})_n$ form an orthogonal basis while $(f_{n,2})_n$ are not orthogonal but still complete (in the limit of $N\rightarrow \infty)$ \cite{fox1994oblique}.  Finally, for a propagating wave incident from medium (1) with angle $\theta$, we have $k_y = k_1^0\sin\theta$.
 
 The boundary condition requires continuity of $\phi$ and $\partial_x\phi$ at $x=0$. We write then:
 \begin{equation} \begin{split}
 \sum_n f_{n,1}(\alpha_{n,1}+\beta_{n,1}) & =  \sum_n f_{n,2}(\alpha_{n,2}+\beta_{n,2})\\
  \sum_n f_{n,1}(\alpha_{n,1}-\beta_{n,1})ik_{x,1}^n  & =  \sum_n f_{n,2}(\alpha_{n,2}-\beta_{n,2})ik_{x,2}^n
\end{split} \end{equation}

By projecting the previous equations on $f_{n,1}$, we can deduce:
 \begin{equation} \begin{split}
2ik_{x,1}^n\alpha_{n,1} & =  \sum_m  \left[i(k_{x,1}^n+k_{x,2}^m)\alpha_{m,2}\right. \\& \left.+i(k_{x,1}^n-k_{x,2}^m)\beta_{m,2}\right]  \langle f_{n,1}, f_{m,2} \rangle \\
2ik_{x,1}^n\beta_{n,1} & =  \sum_m \left[ i(k_{x,1}^n-k_{x,2}^m)\alpha_{m,2} \right. \\& \left. +i(k_{x,1}^n+k_{x,2}^m)\beta_{m,2} \right] \langle f_{n,1}, f_{m,2} \rangle
\end{split} \end{equation}
  
 This allows us to define the transfer matrix as $X_1 = M_{12}X_2$ where $X_i = (\alpha_{0,i},\alpha_{1,i},...,\beta_{0,i},...)$. $M_{21}$ is subsequently defined as $M_{12}^{-1}$.
 
 We finally write the global transfer matrix as $M=M_{12}M_pM_{21}$, where $M_p$ is a diagonal matrix that propagates the modes along the membrane and that is defined as: \begin{equation}\begin{split}M_{p,(n,n)}& = e^{ik_{x,2}^nw}, \\ M_{p,(n+N+1,n+N+1)}& = e^{-ik_{x,2}^nw}, \; \; 0\leq n \leq N \end{split}\end{equation}
 
 We can now write $X_{out} = M X_{in}$ where $X_{in} = (I,R) = (1,0,...,r,\beta_{1,1}...)$ and $X_{out} = (T,0) = (t,\alpha_{1,1},...,0,....)$. By writing $M = \begin{bmatrix} M_1 && M_2 \\ M_3 && M_4\end{bmatrix}$, we have: \begin{equation} T =M_1I+M_2R, \; \; \; \; 0=M_3I+M_4R \end{equation}which allows us to compute the transmission and reflection coefficients as: \begin{equation} R=-M_4^{-1}M_3I, \; \; \; \; T = M_1I+M_2R \end{equation}
 We check of course that $|t|^2+|r|^2=1$.
 
\section{Scattering particles embedded in low-absorbing layer}
  
We consider scattering particles embedded in a layer of index $n$ and negligible absorption in the presence of perfect back-reflector ($R_2=1$). In the limit of large scattering we obtain:  \begin{equation}q(\theta) = 3\cos^2\theta+\frac{2}{\alpha_1}\cos\theta  \end{equation}
where $\theta$ is the \emph{refraction} angle ($<\theta_c=\asin\frac1 n$) and $\alpha_1^{-1} = n^{2}\left[1+\left(1+\frac{1}{n^2}\right)^{\frac 3 2} \right]$.

For isotopic incidence ($f = \frac{n^2}{\pi} \delta(\theta<\theta_c)$), we have: \begin{equation} \langle q \rangle  = \int_{4\pi}q(\theta)f(\theta)d\Omega = 2\pi\frac{n^2}{\pi}\displaystyle\int_{0}^{\theta_c} q(\theta)\sin\theta d\theta = 4n^2\end{equation}

In the presence of bulk scattering, the Yablonovitch limit is indeed maintained for isotropic incidence but can be overcome at normal incidence. 

\bibliography{Ocean_buoy_paper}
\end{document}